# Role of particles size on the cohesive strength of non-sintered (green) ceramics

Authors: M. Hristova,[1] I. Lesov,[1] L. Mihaylov,[2] N. Denkov, S. Tcholakova[1*]

[1]*Department of Chemical and Pharmaceutical Engineering*
[2]*Department of Applied Inorganic Chemistry*
*Faculty of Chemistry and Pharmacy, Sofia University*
*1 J. Bourchier Ave.,1164 Sofia, Bulgaria*

*Corresponding author:
Prof. Slavka Tcholakova
Department of Chemical and Pharmaceutical Engineering
Faculty of Chemistry and Pharmacy, Sofia University
1 James Bourchier Ave., 1164 Sofia
Bulgaria

Phone: (+359-2) 962 5310
Fax:   (+359-2) 962 5643
E-mail: SC@LCPE.UNI-SOFIA.BG

Other authors:
Monika Hristova            *email*: mk@lcpe.uni-sofia.bg
Ivan Lesov                 *email*: lesov@lcpe.uni-sofia.bg
Lyuben Mihaylov            *email*: nhtlm@chem.uni-sofia.bg
Nikolai Denkov             *email*: nd@lcpe.uni-sofia.bg

**ABSTRACT**

Preparation of particle-loaded foams, followed by drying, sintering and/or cross-linking are widely explored routes for developing lightweight ceramics with high mechanical strength. The non-sintered dry ceramic foams are less studied due to their intricate production and the assumed poor mechanical strength of the obtained "green" materials. Here we produce lightweight ceramics from foamed particle suspensions containing spherical silica particles with radii varied between 4.5 nm and 7 μm. The wet foams were prepared in the presence of cationic surfactant and were dried at ambient conditions to obtain porous materials with mass densities between 100 and 700 kg/m$^3$. The materials containing smaller particles exhibited much higher strength (up to 2000 times), approaching that of the sintered materials. A new theoretical expression for predicting the mechanical strength of such materials was derived and was used to explain the measured strengths of the produced materials through the van der Waals attractions between the particles in the final dry materials.

1. Introduction

The development of lightweight porous materials has been of vivid interest to the ceramic community [1-6]. Numerous methods for foam generation and modifications were proposed to deliver lightweight materials with sufficiently high mechanical strength [1-10]. Procedures of gel-casting [1-3], freeze-drying [4], sacrificial replication [5], and supercritical gel-drying [6] were developed in combination with several physical or chemical treatments, such as sintering or chemical cross-linking, to deliver the ultimate performance of these materials. As a result, novel and functional materials with relatively low mass density and high mechanical strength were designed from particles of alumina, silica, silicon carbide, zirconia, hydroxyapatite, graphene oxide, and many others [7-10]. Outstanding as they are, these materials require elaborate, expensive, and energy-demanding manufacturing processes.

The direct foaming technique, combined with ambient drying, is the cheapest, greenest, and least energy-demanding technique for porous ceramics fabrication [11]. However, it is often neglected for two main reasons: elaborate development of the materials (difficult foam generation and foam stabilization, cracking and shrinking upon foam drying) and expected poor mechanical strength of the final porous materials. For example, Gonzenbach et al. [12-13] used different sizes of alumina particles ranging from 28 to 1800 nm to prepare alumina foams which were sintered to increase their mechanical strength. The authors found a minor decrease of the suspension foaminess from 90 to 80 vol% air upon increasing the particle size without explicitly clarifying the effect of particle size on the mechanical properties of the final materials.

A general trend of obtaining higher flexural strength when using smaller particles upon sintering dense ceramics was reported for $B_4C$ and SiC materials [14-15]. However, little to no experimental evidence was given for non-sintered materials, apart from the experiments with thin colloidal films, where the film hardness increases with the decrease of particles size [16]. Therefore, we decided to perform a systematic study on the effect of particles size on the cohesive strength of non-sintered porous materials obtained from foamed suspensions of spherical silica particles.

To perform such a systematic study, we applied the direct foaming method on silica suspensions containing a cationic surfactant, followed by ambient drying of the wet foam. We varied the size of the spherical silica particles and the amount of trapped air to produce materials with different levels of porosity. The cationic surfactant hydrophobized the surface of the silica particles, thus driving the particles to adsorb on the bubble surfaces and to attract each other in the bulk phase via hydrophobic attraction [17]. The tuned particle interactions allowed us to produce ceramic materials with various mass densities.

We produced porous materials with mass densities varied between 100 and 700 kg/m$^3$. Interestingly, the materials obtained from smaller particles exhibited relatively high mechanical strength, approaching that of sintered materials with similar characteristics [12-13]. This higher strength was explained by considering the van der Waals attraction between the solid particles in the final porous materials. The provided explanations allow one to design stronger green materials by selecting appropriate particle morphology and surface modifications, as well as better understanding of the structure-strength relationship during partial sintering of ceramics with different particle sizes.

## 2. Theoretical background

The mechanical strength of non-sintered porous materials is expected to depend on the particle size. This strength may be governed by the attractive forces acting between the particles, with possible bridging of the particle surfaces via hydrogen bonds and/or even via possible formation of covalent bonds, created through the condensation of surface -OH groups located on different particles (in the zones of their direct contact). Conceptually, the porous green materials are close-packed assemblies of ceramic particles of a given size, with included pores between them. These pores could originate from the entrapment of bubbles during the foam generation, from the interparticle space upon close particle packing, and from micro-cracks formed during foam drying. Experimental measurement of the foam mass during drying shows that virtually all water is evaporated, which causes the materials to shrink significantly.

In the following, we estimate the role of attractive forces between the contacting particles and check whether van der Waals forces are dominant or we should also account for other additional forces to explain the particle binding in these materials. With this aim in view, hereby, we estimate the expected order of magnitude of van der Waals forces based on the particle size and material porosity.

The Gibson-Ashby model implies that the mechanical strength of the porous material, $\sigma_{Cr}$, depends on the mechanical strength of the wall matrix, $\sigma_{wall}$, the mass density of the wall matrix, $\rho_{wall}$, and the mass density of the porous material, $\rho$ [18]:

$$\frac{\sigma_{Cr}}{\sigma_{wall}} = f\left(\frac{\rho}{\rho_{wall}}\right) \qquad (1)$$

where $f(x)$ is a function which depends on the detailed structure of the porous material. The functional dependence of $f$ on the structure of porous material is given as [18]:

$$f\left(\frac{\rho}{\rho_{wall}}\right) = \left(\frac{\rho}{\rho_{wall}}\right) \qquad \text{(closed cells)} \quad (2a)$$

$$f\left(\frac{\rho}{\rho_{wall}}\right) = 0.2\left(\frac{\rho}{\rho_{wall}}\right)^{1.5} \qquad \text{(open cells)} \quad (2b)$$

In our previous study [23], we showed experimentally that both silica and carbonate materials showed a smooth transition from the Gibson-Ashby theoretical predictions for closed cell structure (at high mass density) toward the theoretical prediction for open cell structure (at low mass density). For the data in the transition region it is shown in Ref. [17] that the mechanical strength of the porous materials prepared from silica-TTAB suspensions having $\rho_{wall}$ = 474 kg/m$^3$ and $\sigma_{wall}$ = 1463 kPa can be described as $\sigma_{Cr} = 1.9\rho^{2.2}$. Therefore the expression derived in [17] can be re-written as $\frac{\sigma_{Cr}}{\sigma_{wall}} = \left(\frac{\rho}{\rho_{wall}}\right)^{2.2}$ and $f$ can be expressed as:

$$f\left(\frac{\rho}{\rho_{wall}}\right) = \left(\frac{\rho}{\rho_{wall}}\right)^{2.2} \quad \text{(transition between open and closed cells)} \quad (2c)$$

In another publication [19], we established that the density of the porous ceramics is linearly dependent on the bubble air volume fraction in the wet foam, $\Phi_0$, and on the mass density), $\rho_{wall}$, of the closely-packed particles in the dry walls of the porous materials (around the voids left by the bubbles):

$$\rho = \rho_{wall}(1-\Phi_0) \quad (3)$$

The combination of Eqs. [1-3] leads to the following relation for the dry porous materials obtained from foamed suspensions of solid particles:

$$\frac{\sigma_{Cr}}{\sigma_{wall}} = (1-\Phi_0) \quad \text{(closed cells)} \quad (4a)$$

$$\frac{\sigma_{Cr}}{\sigma_{wall}} = 0.2(1-\Phi_0)^{1.5} \quad \text{(open cells)} \quad (4b)$$

$$\frac{\sigma_{Cr}}{\sigma_{wall}} = (1-\Phi_0)^{2.2} \quad \text{(transition between open and closed cells)} \quad (4c)$$

In the following consideration, we estimate the mechanical strength of the wall matrix, considering the forces between the neighboring particles in the dry wall. Assuming that the particle matrix is compressed by solid wall, as shown in Figure 1, the wall destruction would occur when the external force overcomes the attraction between the neighboring particles. Based on this idea, one can define the strength of the dry matrix of particles as:

$$\sigma_{wall} = \frac{F_W}{S_1} \quad (5)$$

where $F_W$ is the force acting on the wall and $S_1$ is the area from the wall upon which this force is applied. Considering one particle at the matrix boundary, which is in contact with the solid wall, the force acting between the particle and the wall is counterbalanced by the vertical projection of the forces, $F_p$, of the same particles with its neighbors from the second layer toward the matrix interior, see Figure 1, which is given by the equation [20]:

$$F_W = nF_p \cos\theta \quad (6)$$

Where $n$ is the number of close neighbors located in the second layer, $\theta$ is the angle between the direction of the particle-particle forces and the normal force acting on the solid wall. The values of $n$ and $\theta$ depend on the type of packing. For fcc-packing of the particles, the values are: $n = 3$ and $\cos\theta = 0.816$ [20]. The cross-section of the area per one particle on the wall also depends on the packing and can be expressed as:

$$S_1 = \frac{\pi R_p^2}{\varphi_p} \quad (7)$$

Where $\varphi_P$ is the particle close-packing fraction in two dimensions, which is e.g. 0.91 for hexagonal packing or 0.78 for square (body-centered cubic) packing [21].

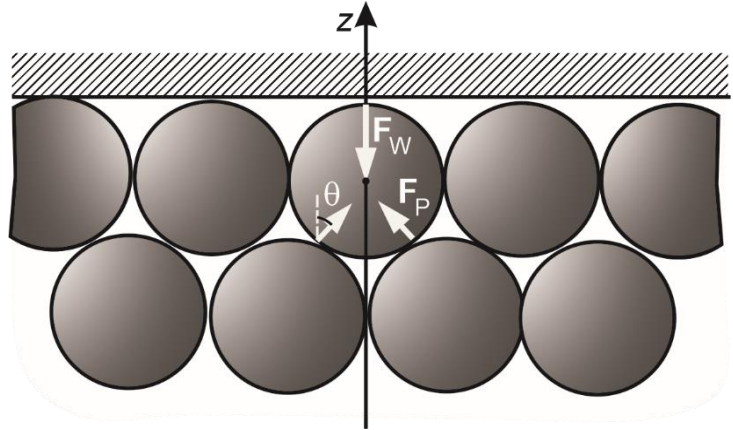

**Figure 1.** Schematic representation of the particles in contact with a solid wall.

Combining eqs. (5-7) we obtain the following expression for the mechanical strength of the porous material:

$$\sigma_{wall} = F_p \frac{n\varphi_p \cos\theta}{\pi R_p^2} \tag{8}$$

Assuming that the prevailing attractive force between the particles is van der Waals (vdW), $F_p$ can be written as [22]:

$$F_p = \frac{A_H R_P}{12h^2} \tag{9}$$

Here $A_H = 6.3\times10^{-20}$ J is the Hamaker constant for $SiO_2$ in air [22], $R_p$ is the average particle radius, and $h$ is the distance between two neighboring interacting particles.

Therefore, the mechanical strength of a porous material, held by van der Waals forces between the particles, is expressed as:

$$\sigma_{wall} = \frac{nA_H \cos\theta}{12\pi R_p h^2}\varphi_p \tag{10}$$

For fcc packing n = 3; $\cos\theta = 0.816$; $\varphi_P = 0.91$. For silica particles interacting through the air $A_H = 6.3\times10^{-20}$ J. Here $R_p$ is the particle radius and one can deduce from eqs. (8) and (9) that the mean surface-length radius, $R_{p21}$, should be used for polydisperse particles.

$$\sigma_{Cr} = \frac{nA_H \cos\theta}{12\pi R_p h^2}\varphi_p(1-\Phi_0) = \frac{nA_H \cos\theta}{12\pi R_p h^2}\varphi_p \frac{\rho}{\rho_{wall}} \quad \text{(closed cells)} \tag{11a}$$

$$\sigma_{Cr} = 0.2\frac{nA_H \cos\theta}{12\pi R_p h^2}\varphi_p(1-\Phi_0)^{1.5} = 0.2\frac{nA_H \cos\theta}{12\pi R_p h^2}\varphi_p \left(\frac{\rho}{\rho_{wall}}\right)^{1.5} \quad \text{(open cells)} \tag{11b}$$

$$\sigma_{Cr} = \frac{nA_H \cos\theta}{12\pi R_p h^2}\varphi_p(1-\Phi_0)^{2.2} = \frac{nA_H \cos\theta}{12\pi R_p h^2}\varphi_p \left(\frac{\rho}{\rho_{wall}}\right)^{2.2}$$

(transition between open and closed cells)(11c)

As explained above, $\Phi_0$ in eq. (11) is the air volume fraction in the wet precursor foam. To test the validity of eqs. (11a-11c) we prepared a series of porous materials from particle suspensions with radius varied between 7 nm and 5 μm from wet foams having an initial air volume fraction $\Phi_0$ varied between 26 % and 76 %. Note that the drying of particle suspension without incorporated air in it led to cracking of the material during drying, therefore, we could not measure directly the mechanical strength of the close-packed material, $\sigma_{wall}$.

## 3. Methods and materials

### 3.1 Materials

We used five types of spherical silica particles – three Ludox suspensions with particle size between 8 and 25 nm and two samples of Excelica with mean sizes of 3 and 30 μm according to their producers. The particle properties according to the producers are shown in Table S1. We used tetradecyl trimethyl ammonium bromide, TTAB, as cationic surfactant (>99%, Sigma-Aldrich) to modify the particle surface and obtain Pickering foams with small bubbles which do not coarsen with time. The pH of the silica suspensions was adjusted to pH=8.5 using solutions of NaOH (≥ 99.9%, Sigma-Aldrich) or HCl (37%, Sigma-Aldrich). Deionized water from Elix 5 module (Millipore Inc., USA) was used to prepare all solutions.

### 3.2. Methods

**Suspension preparation.** Colloidal suspensions from Excelica particles were prepared by dispersing the desired amount of silica powder in deionized water and adjusting the pH to 8.5. Afterward, the suspensions were sonicated for 10 min at 1200 W using an Ultrasonic pulse homogenizer (SKL1500-IIDN, Ningbo Sklon Lab Instrument Co. Ltd). We used Ludox particles as received (30 wt% colloidal suspensions) or concentrated them via rotatory vacuum evaporator (R-210, Buchi). After the concentration, we adjusted the suspension pH and sonicated the suspensions for another 10 min. Suspensions were left to cool down after the sonication to room temperature before using them in the following experiments.

**Foam preparation and drying.** We foamed 400 g suspensions using the desired amount of TTAB (25 wt% stock solution). Immediately after the surfactant was added, we mixed the suspensions with Kenwood Chef Premier KMC 560, 1000 W. The foaming continued until we reached the desired air entrapment determined gravimetrically. After the foams were generated, they were placed in 114 mL cylindrical Teflon molds with 7.5 cm diameter and 2.6 cm height and let to dry under ambient temperature and humidity (25 ± 3°C and 45 ± 10 % RH).

Assuming complete surfactant adsorption on the particle interface, as in Ref. 17, we calculated the adsorption, $\Gamma$ in μg/m$^2$, using the equation:

$$\Gamma = C_s/(AC_p) \cdot 10^6 \qquad (12)$$

Where $C_s$ and $C_p$ are the surfactant and particle concentrations in wt%, while $A$ is the specific surface area in m$^2$/g of the particles, see Table S1.

**The mass density** of the obtained porous materials was determined by measuring the mass of the dry material and dividing it by its dry volume.

**The mechanical strength** of the final porous materials was determined using a universal testing machine Tira GmbH Tiratest 2300. The stress was measured with ±1% accuracy, as a function of the deformation, at rates varied between 0.5 and 5 mm/min. The samples had sizes between 50 and 70 mm in diameter and around 1.5–2.5 cm in height. The crushing stress of these materials was determined from the materials yielding point (see Figure S1). The materials with low mechanical strength, e.g., below 40 kPa, easily deformed upon manual handling. We used a stainless-steel ball to measure their strength, which was placed on top of the porous materials. We measured the ball imprint on the surface and calculated the yield stress/crushing strength, as explained in Ref. [23]. All measurements were performed at ambient conditions: 25 °C and 45 ± 10% RH.

**SEM.** Table-top SEM Hitachi TM4000 was used to visualize the structure of the final ceramic porous materials. The dry foams were manually fractured and observed at different magnifications.

**TEM.** High-Resolution Transmission Electron Microscope (JEM2100, JEOL) was used for studying the size of the nanoparticles.

## 4. Experimental results

The particle size distribution as measured from SEM and TEM images of the used particles are shown in Figure 2 and Figure S2. The nanometer particles are monodisperse and the experimental data are fitted by normal distribution, whereas the micrometer particles are very polydisperse and log-normal distribution is used. The obtained results for the most probable radius, $R_m$ and mean area-length radius $R_{21}$ are shown in Table 1. Note that for polydisperse samples the average particle radius that should be substituted in eq. (11a-11c) is the mean area-length radius, $R_{21}$. As can be seen, the results for the nanometer particles are in a good agreement with the results provided by the producer. However, due to the polydispersity of the micrometer particles, the most probable particle size and the mean area-length radius are significantly smaller than the particle size provided by the producer for Excelica SE40 samples.

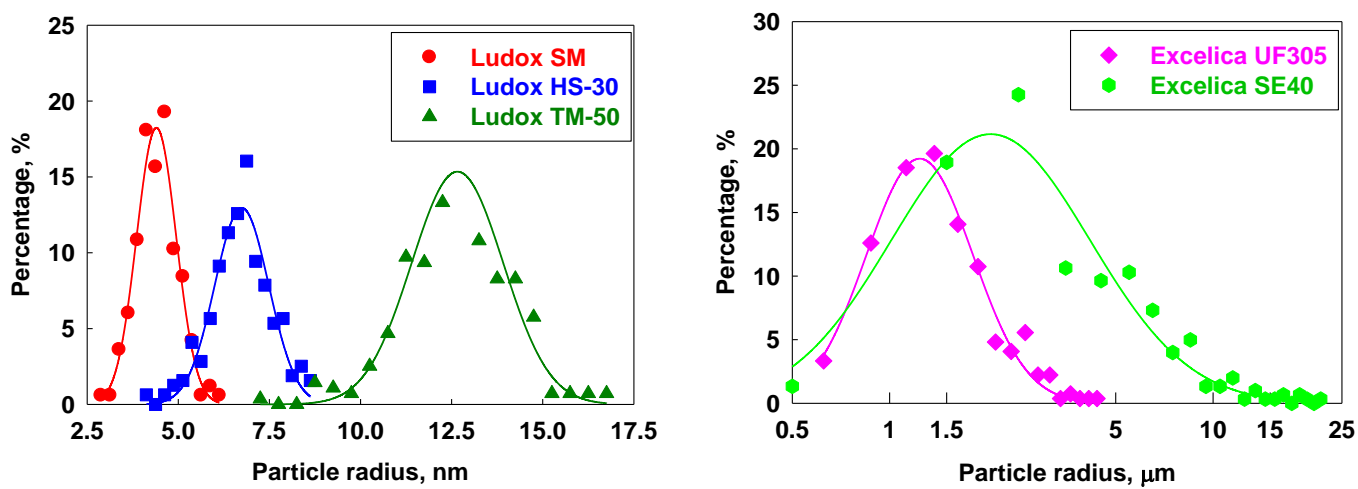

**Figure 2.** Particle size distribution by number for (A) nanometer (measured via TEM) and (B) micrometer particles (measured via SEM). The points are experimental data, whereas the curves are the best fits to these data by (A) normal distribution and (B) log-normal distribution.

**Table 1.** The properties of used particles and abbreviation used in the text.

| Trade name | Abbreviation used in the text | Initial pH | Specific surface area, $A$, $m^2/g$ from producer | Mean radius from producer, nm | $R_m$, nm | $R_{21}$, nm |
|---|---|---|---|---|---|---|
| Ludox SM | L-5 | 10.0 | 340 | 3.5 | 4.4±0.6 | 4.5±0.6 |
| Ludox HS-30 | L-7 | 9.5 | 220 | 6.0 | 6.8±0.8 | 6.8±0.8 |
| Ludox TM-50 | L-13 | 9.0 | 140 | 11.0 | 12.7±1.3 | 12.8±1.3 |
| Excelica UF305 | E-2 | 7.0 | 2.1 | 1350 | 1240 | 1880 |
| Excelica SE40 | E-7 | 7.0 | 0.6 | 19000 | 2050 | 7210 |

Based on our earlier experience with fractal silica particles, we first tried to foam 30 wt% silica suspensions adding TTAB into the suspension to ensure 30 μg/m² surface coverage [17]. However, we found that either no air was trapped under these conditions (when using micrometer particles) or the foams produced from suspensions of nanoparticles (though being of 75-80 vol.% air initially) were unstable with time and suffered from slow drainage of the suspension from the foam. As explained in our previous studies [17,23-24], such drainage leads to inhomogeneous distribution of water and particles in the drying foams with subsequent severe cracking of the porous materials in the final stages of their drying.

Therefore, we gradually increased the surfactant concentration until the particles in the foams started to aggregate under the action of hydrophobic forces [24-26] and sufficiently high yield stress of the suspension was generated to stabilize the foams against drainage [26-27]. The larger micro-particles required 3 to 4 times higher surfactant adsorption to stabilize the foams against drainage, as illustrated in Figure 3.

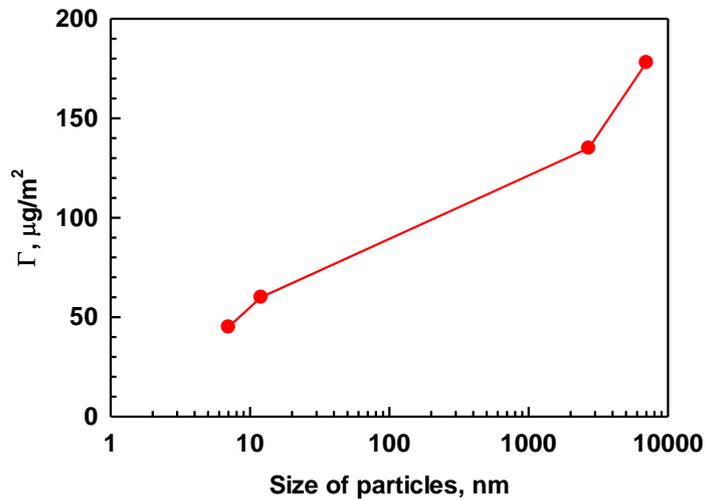

**Figure 3.** Minimum TTAB adsorption needed to stabilize against liquid drainage the foams produced from 30 wt% silica suspensions, as a function of the particles' radius given by the producer.

To investigate the role of the particle concentration and in an attempt to reduce the material shrinking upon drying [19,23], we increased the particle concentration in the foamed suspensions up to 50 wt % or 60 wt%, as listed in Table 2. Samples from the respective foamed suspensions, with different air volume fractions, were left to dry at ambient temperature and humidity for several weeks until they reached a constant weight. After drying, we recorded each sample's weight and volume and calculated their green densities, summarized in Table 2. Figure 4 shows the dry ceramic foams' mass density as a function of the initial bubble air volume fraction for all materials obtained. As one can see, there is a good correlation between the initial bubble volume fraction and the final mass density of the dried materials, as anticipated by eq. (3), see Fig. 4 and Figure S3 in supporting information. The estimated wall mass density is $\rho_{wall} \approx 1000$ kg/m$^3$ for all Ludox particles and for Excelica 305, whereas slightly higher value is determined for the materials prepared from Excelica SE40 of $\rho_{wall} \approx 1200$ kg/m$^3$. From the wall mass density, the particle fraction in the wall is estimated to be $\approx 0.45$ for Ludox particles and for Excelica 305 and $\approx 0.53$ for Excelica SE40. These particle fractions are much smaller as compared to the particle fractions corresponding to fcc close packing which is $\approx 0.74$ [21]. Therefore, we could expect that our materials would have somewhat lower mechanical strength as compared to one predicted on the base of fcc packing.

**Table 2.** Experimental conditions for preparing porous materials from particle-stabilized foams, and their corresponding dry densities.

| Commercial particles | $C_p$, wt% | $\Gamma$, μg/m$^2$ | $\Phi_0$,% | $\rho$, kg/m$^3$ |
|---|---|---|---|---|
| Ludox SM (L-5) | 30 | 46 | 45 | 547 |
| | 30 | 61 | 40 | 639 |
| | 30 | 60 | 65 | 344 |
| | 30 | 60 | 65 | 345 |
| | 38 | 30 | 56 | 530 |
| | 40 | 30 | 62 | 379 |
| Ludox HS30 (L-7) | 30 | 60 | 85 | 127 |
| | 30 | 60 | 76 | 207 |
| | 30 | 93 | 60 | 381 |
| | 30 | 75 | 64 | 338 |
| Ludox TM50 (L-13) | 60 | 18 | 49 | 639 |
| | 50 | 45 | 64 | 349 |
| | 50 | 45 | 70 | 299 |
| | 30 | 30 | 63 | 300 |
| | 30 | 30 | 74 | 195 |
| | 30 | 30 | 77 | 162 |
| | 60 | 60 | 49 | 639 |
| | 50 | 50 | 64 | 349 |
| | 50 | 50 | 70 | 299 |
| | 50 | 33 | 73 | 272 |
| | 50 | 33 | 76 | 252 |
| | 50 | 33 | 78 | 212 |
| | 50 | 33 | 71 | 262 |
| | 50 | 33 | 77 | 236 |
| Excelica UF305 (E-2) | 60 | 45 | 26 | 792 |
| | 60 | 45 | 47 | 577 |
| | 30 | 135 | 76 | 114 |
| | 30 | 225 | 76 | 117 |
| Excelica SE40 (E-7) | 60 | 135 | 42 | 722 |
| | 60 | 253 | 55 | 511 |
| | 60 | 312 | 63 | 466 |

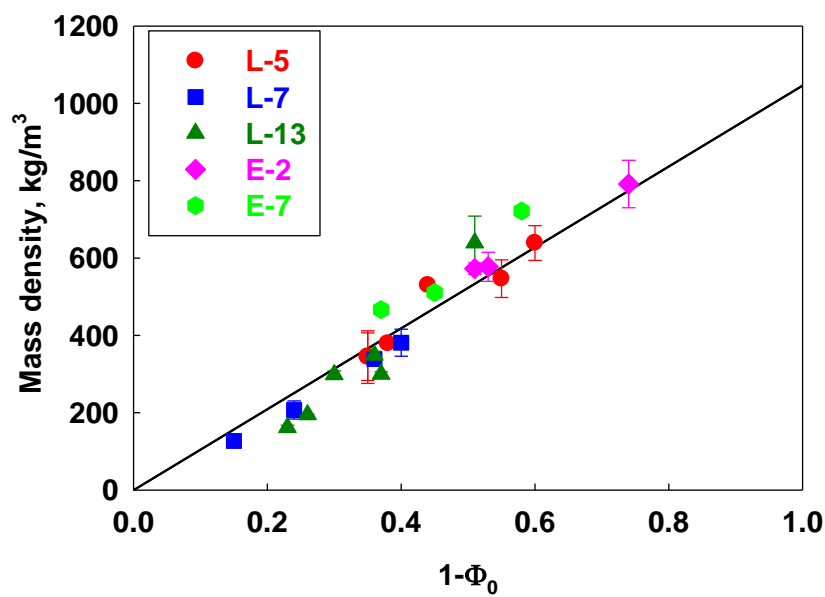

**Figure 4.** Mass density of porous materials as a function of volume fraction of suspension for materials prepared with particles having different sizes.

We measured the mechanical strength of the final dry materials using the procedure described in section 3.2 and plotted their compressive strength as a function of the material mass density, see Figure 5. The mechanical strength of the green materials, prepared with Ludox SM ($R_{21}$ = 4.5 nm) silica particles, was exceptionally high – more than 4 MPa strength which is a value comparable to the sintered $Al_2O_3$ materials obtained by Gonzenbach et al. [7] at 24 % relative density. Unfortunately, the increased mechanical strength of these materials was accompanied with decreased mechanical integrity upon drying. Foams produced from Ludox SM particles ($R_{21}$ = 4.5 nm) suffered from tremendous capillary pressures (proportional to the inverse particle radius), and multiple cracks were formed while drying. Materials with lower mass density suffered more cracks as seen in Figure 5B. Therefore, despite their high mechanical strength, the materials obtained with smaller particles could not be dried as one large piece of centimeter size under these conditions – instead, multiple pieces with high mechanical strength were formed.

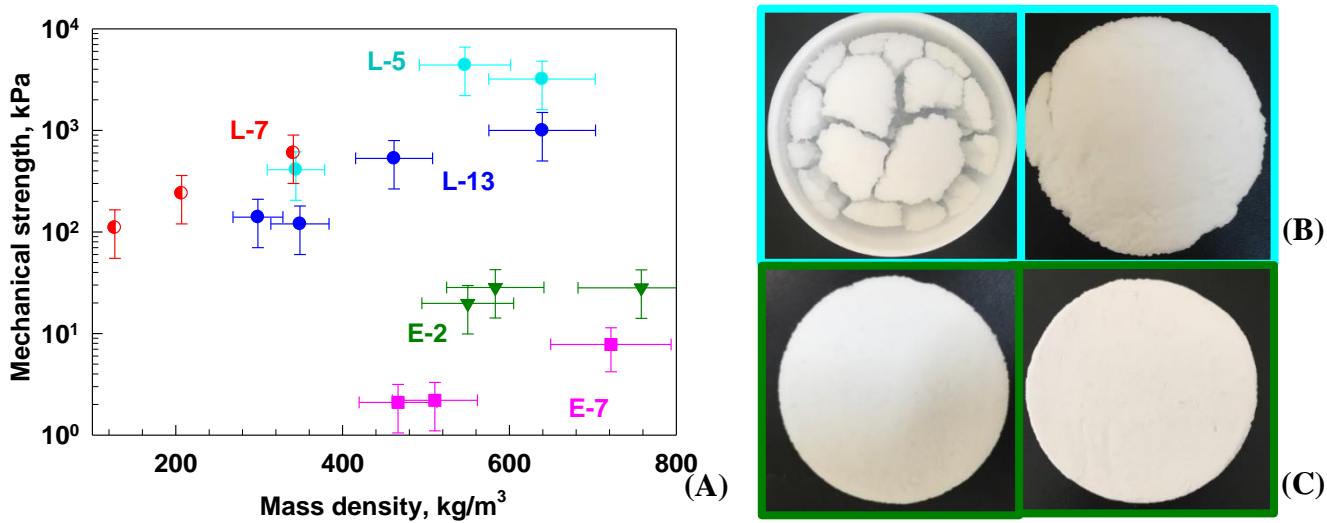

**Figure 5** (A) Compressive strength of the dry materials versus their mass density, for particles with different sizes used for the foam preparation. (B, C) Macroscopic images of materials prepared from (B) Ludox SM and (C) Excelica UF305. The images on the left correspond to $120 \pm 20$ kg/m$^3$, while those on the right are $600 \pm 60$ kg/m$^3$.

Increasing the particle size from $R_{21} = 4.5$ nm to $R_{21} = 6.8$ nm and 12.8 nm led to a significant decrease in the measured mechanical strength at the same mass density of the dry materials without significantly improving the mechanical integrity of the foams.

The transition from nanometer to micrometer particles led to a qualitative change in the materials' behavior. Materials from Excelica particles with $R_{21} = 1.9$ μm particles dried without formation of macroscopic cracks at all tested air volume fractions (see Fig. 5C). However, the materials had 10-100 times lower mechanical strength as compared to their nanometric counterparts. The mechanical strength of the materials with the largest particles was exceptionally low. Upon mild touch, the materials obtained with 7.2 μm in size particles broke into pieces and had few kPa strength only.

The microstructure of the materials with the lowest mass density is illustrated in Figure 6A,B with micrographs obtained by SEM at different magnifications. The material prepared from Ludox HS with $R_{21} = 6.8$ nm particles had around 10 times smaller bubbles/macropores compared to the 1.9 μm particle-containing material, at the same concentration of particles. Although the foam films of the 6.8 nm material were mildly wrinkled, the ratio between the Plateau channel cross-section and the foam film thickness was largely in favor of the Plateau channels, implying that the breakage of the porous material should occur via fracturing of the channels for both foams, even in those having high air volume fractions.

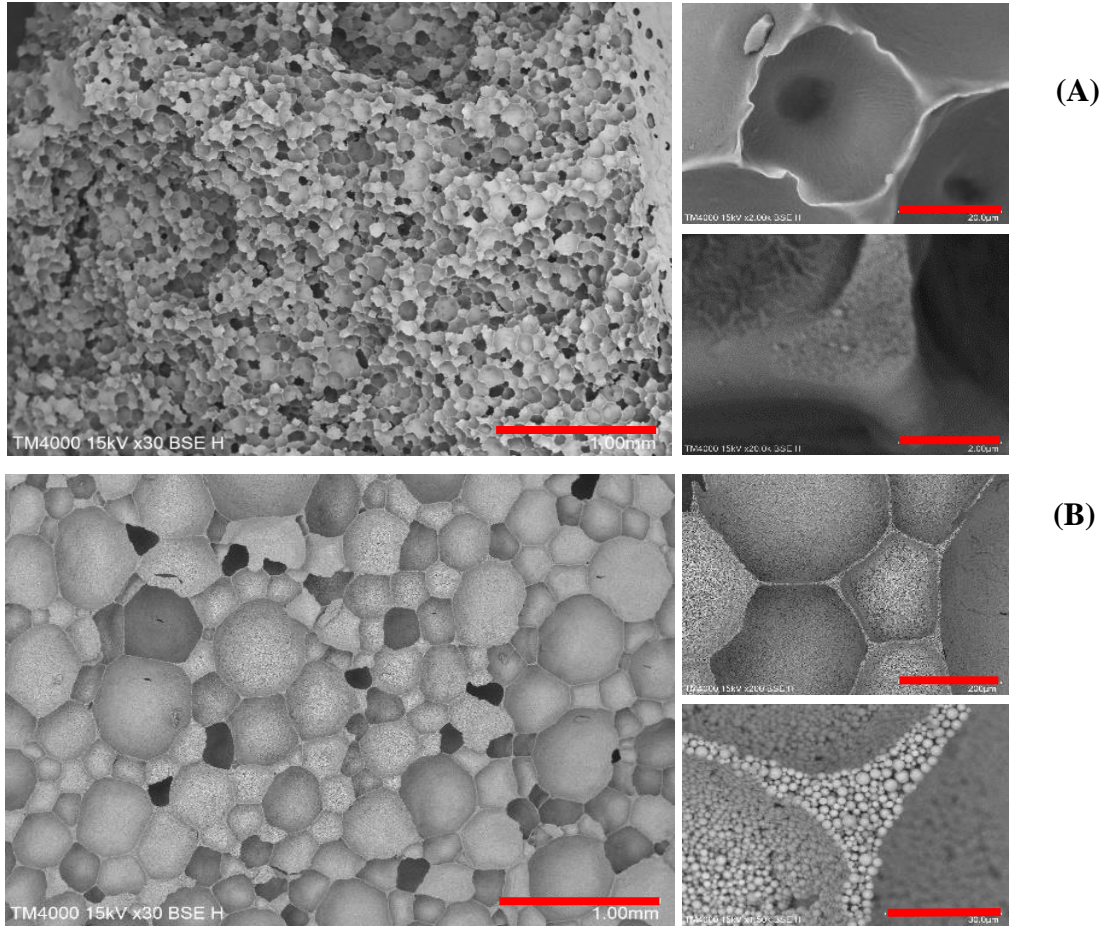

**Figure 6**. Microstructure of the porous materials with 76 vol% bubbles, prepared from (A) Ludox TM with $R_{21}$ = 12.8 nm precipitated silica particles and **(B)** Excelica 305 with $R_{21}$ = 1.9 µm fused silica particles at different magnifications. Scales are 1000 µm, 200 µm and 30 µm.

**Discussion**

To test the validity of eqs. (11a-11c), we have to define the thickness, $h$, between the particles at which the compressive strength is measured. To overcome this problem, the experimental data shown in Figure 5 are replotted in Figure 7A in the scale $\ln(\sigma_{CR}R_p)$ vs $\ln(\rho/\rho_{wall})$. One sees that the slope of the respective line is very close to 2.2, which confirms that eq. (11c) can be used to describe the experimental data. The latter conclusion implies that the prepared materials fall within the transition window between the closed and opened cells. It is also seen that the results for Ludox particles fall on a master line after accounting for the size of these particles and the same is true for the materials prepared from Excelica particles, while the two master lines are parallel to each other. From the intercepts of these lines we determined the values of $\sigma_{wall}R_p = \dfrac{nA_H \cos\theta}{12\pi h^2}\varphi_p$ = 40 ± 20 mPa.m for Ludox particles and 140 ± 35 mPa.m for Excelica particles. assuming that the particles are organized in fcc-packing which (as discussed above) is not strictly valid, the estimated thickness of the particle-particle contact is $h \approx 0.30\pm0.08$ nm for Ludox and $h \approx 0.16\pm0.02$ nm for Excelica particles. Note that the latest value is very close to the cut-off separation proposed by Israelachvili [22] for determination of the adhesion energy between contacting surfaces which was found to be ≈ 0.165 nm [22].

The smaller value of *h* determined for Excelica type of particles is most probably due to the polydispersity of the particles and the presence of smaller particles which are not well visible via the SEM microscope at the magnification used – the actual particle size could be smaller than the size presented in Table 1. This hypothesis is supported by the higher adsorption of TTAB required to obtain stable porous material as can be seen from Figure 2, which is ≈ 3-times higher as compared to the adsorption on Ludox particles.

Another possibility is the difference in the surface Si-OH groups density. Ludox particles are precipitated silica with approximately 4.6 Si-OH groups per $nm^2$, while Excelica are fused silica produced by vapor phase reaction that has large part of these-OH groups reduced to Si-O-Si groups at the high processing temperature [28]. Given the bond length of Si-OH is around 0.165 nm [29,30], precipitated silica intraparticle distance corresponds to around two surface groups distance. On the other hand, fused silica shows distance of either one Si-OH group or about two Si-O-Si group perpendicular to the particle surface [31]. Either way such difference in the separation distances were recently observed for hydrophilic and fused silica by the means of a surface-force apparatus [30].

By using the estimated values of the thickness in the point of contact between the particles, we calculated the theoretical stress for the materials formed from different particles and compared it with the experimentally determined one – a very good agreement between theory and experiment is seen in Figure 7.

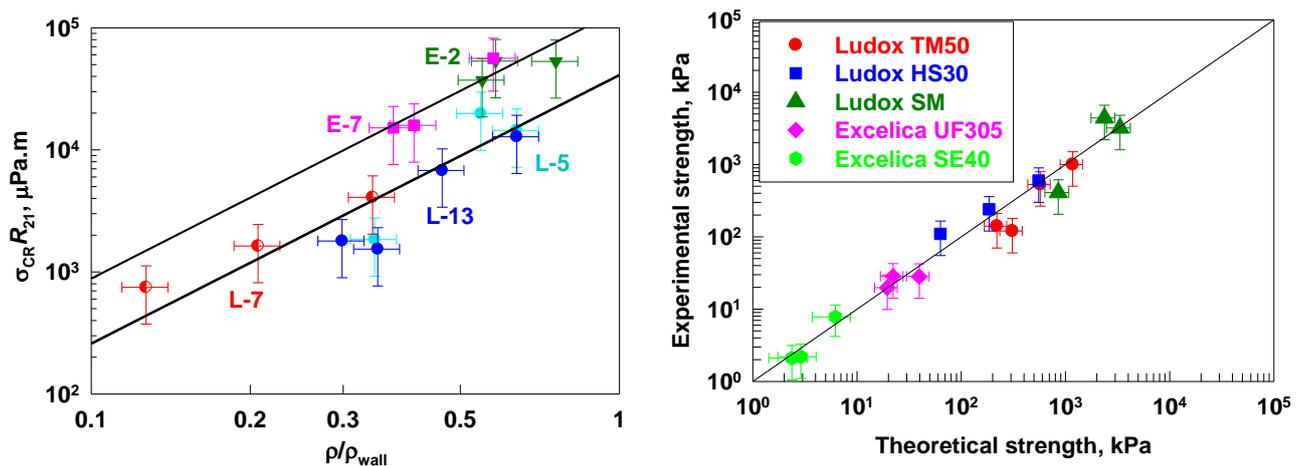

**Figure 7.** **(A)** Experimental data from Figure 5 plotted in the scale $\ln(\sigma_{CR}R_p)$ vs $\ln(\rho/\rho_{wall})$. **(B)** Experimentally measured strength for all porous materials studied here, as a function of the calculated strength via eq.(11c) for the different silica particles used. The errors for the theoretically estimated strengths reflect the uncertainty in the particle size distribution. The parameters used in the calculations are for fcc packing n = 3; $\cos\theta$ = 0.816; $\varphi_P$ = 0.91; $A_H$ = $6.3\times10^{-20}$ J; $h \approx 0.30$ nm for Ludox and $h \approx 0.163$ nm for Excelica particles.

Although the agreement between the theory and experiment is reasonable, the experiment does not completely exclude the possible formation of hydrogen and/or covalent bonds upon direct particle contact, as suggested in Refs. 32-33]. Lai, Shi and Huang [32] estimated and measured the detachment force for a point contact for silicon AFM tip with a curvature similar to Ludox TM particles to be around 313.9 nN for Si-O-Si bonds. Applying this value for closely-packed 25.3 nm Ludox TM particles, we estimate the gigantic 1.80±0.9 GPa separation force for covalent bonds and > 0.25 GPa for hydrogen bonds obtained at 20-50°C and 30-60 % air humidity ($F_{adh}$ > 40 nN). However, we determined experimentally this force to be by three orders of magnitude smaller, ≈ 2.5 MPa. These estimates show that covalent and hydrogen bonding can add an immense value to the mechanical strength of the green ceramic if an efficient post drying chemical or thermal treatment is applied (whereas pre-drying treatment might affect intraparticle distance, $h$). On the other hand, via these estimates we have verified *aposteriori* our assumption that the van der Waals forces between the silica particles in the studied dry materials are most probably the dominant interactions.

Last, we assessed the effect of the capillary bridge forces which may appear due to a small fraction of residual water in the final dry materials. We dried Ludox TM50 materials for 3-5 days at 70°C and then sealed them in plastic bags without air to avoid restoring the 2.8 wt% capillary water (it took 8 hours to restore 0.8 wt% $H_2O$ without sealing). Then, we measured their mechanical strength immediately after taking them out of the plastic bags, showing around 8 % lower mechanical strength in the dried materials than that of the materials stored in open atmosphere. Therefore, the capillary water had a relatively small effect, within the experimental error, for the mechanical strength of our materials. Thus, we can conclude that the van der Waals forces were dominant in the green materials studied here.

## 5. Conclusions

In this study we prepared green ceramic materials from spherical particles with different sizes, varied from 4.5 nm to 7 μm, and measured the mechanical strength of the obtained materials, as a function of their mass density. We established that the mechanical strength of the materials increases significantly with the decrease of particle size. The obtained experimental results are described by a semi-empirical model which accounts for the van der Waals attraction between the silica particles in the dry porous materials. This model could be used to optimize the mechanical strength of the materials in future studies, based on the set of involved parameters, such as the particles size and their chemical nature. The model could be helpful also to assess the effects of possible chemical bonding and/or sintering of the particles in parallel or subsequent surface treatment, or even to allow better understanding of partial sintering in materials, where surface chemistry and morphology change upon heating.

**CRediT author statement**
**MH –** Investigation, Validation, Visualization
**IL –** Investigation, Conceptualization, Methodology, Formal analysis, Visualization, Supervision, Writing - Original Draft
**LM –** Investigation, Resources
**ND –** Conceptualization, Methodology, Writing - Review & Editing
**SC –** Supervision, Conceptualization, Methodology, Formal analysis, Writing - Review & Editing

**Acknowledgements**
The authors are grateful to the Center of Excellence in Mechatronics and Clean Technologies, "Science and Education for Smart Growth" Operational Program 2014-2020 under the procedure for selection of projects BG05M2OP001-1.001 "Construction and development of centers of excellence"
The authors also express their gratitude towards Assoc. Prof. R. Krastev for mechanical testing, the students J.-S. Markusova and T. Arnaudova for some of the foaming experiments and suspension drying experiments, and to V. Yordanova for measuring the particle distributions from SEM and TEM micrographs.

**Declaration of interests**

☒ The authors declare that they have no known competing financial interests or personal relationships that could have appeared to influence the work reported in this paper.

☐ The authors declare the following financial interests/personal relationships which may be considered as potential competing interests: